\documentclass[referee]{aastex631}

\shorttitle{AASTeX v6.3.1 Sample article}
\shortauthors{Guo et al.}
\graphicspath{{./}{figures/}}

\begin{document}

\title{Eclipsing millisecond pulsars with He star companions}

\author{Yunlang Guo}
\affiliation{School of Astronomy and Space Science, Nanjing University, Nanjing 210023, China}
\affiliation{Key Laboratory of Modern Astronomy and Astrophysics, Nanjing University, Ministry of Education, Nanjing 210023, China}

\author{Bo Wang}
\affiliation{Yunnan Observatories, Chinese Academy of Sciences, Kunming 650216, China}
\affiliation{International Centre of Supernovae, Yunnan Key Laboratory, Kunming 650216, China}

\author{Xiangdong Li}
\affiliation{School of Astronomy and Space Science, Nanjing University, Nanjing 210023, China}
\affiliation{Key Laboratory of Modern Astronomy and Astrophysics, Nanjing University, Ministry of Education, Nanjing 210023, China}

\author{Dongdong Liu}
\affiliation{Yunnan Observatories, Chinese Academy of Sciences, Kunming 650216, China}
\affiliation{International Centre of Supernovae, Yunnan Key Laboratory, Kunming 650216, China}

\author{Wenshi Tang}
\affiliation{Department of Astronomy, Xiamen University, Xiamen 361005, China}

\correspondingauthor{Yunlang Guo, Bo Wang, Xiangdong Li}
\email{yunlang@nju.edu.cn, wangbo@ynao.ac.cn, lixd@nju.edu.cn}
\begin{abstract}
Eclipsing millisecond pulsars (MSPs) are a type of pulsar binaries
with close orbits ($\lesssim1.0\,$d).
They are important objects for studying the accretion history of neutron stars (NSs), pulsar winds,
and the origin of isolated MSPs, etc.
Recently,
a new eclipsing MSP, PSR J$1928+1815$,
was discovered by the Five-hundred-meter Aperture Spherical radio Telescope.
It is the first known pulsar with a He star companion,
as suggested in Yang et al.
The system features a short orbital period of $\sim0.15\,$d and a relatively massive companion $\gtrsim1.0\,M_\odot$.
However,
the origin of PSR J$1928+1815$ remains highly uncertain.
In this paper,
we investigated the formation of the new subclass of eclipsing MSPs containing (evolved) He star companions through NS + He star channel.
We found that if a NS binary undergoes subsequent mass-transfer phases following Case BA or Case BB,
it may appear as an eclipsing MSP during the detached phase.
Additionally,
we obtained the initial parameter space for producing eclipsing MSP with He star companions.
Using binary population synthesis approach,
we estimated their birth rate to be $\sim2.1-4.7\times10^{-4}\rm\,yr^{-1}$,
corresponding to a total number of $\sim55-150$ systems in the Galaxy.
Moreover,
we concluded that
PSR J$1928+1815$ may originate from the evolution of an NS+He star system
with an initial orbital period of $\sim0.1$\,d,
which can undergo the Case BB mass transfer.
\end{abstract}

\keywords{Millisecond pulsars (1062) --- X-ray binary stars (1811) --- Close binary stars (254)}

\section{Introduction}
Millisecond pulsars (MSPs), characterized by spin periods of less than $30$ ms,
are believed to form through the evolution of low- and intermediate-mass X-ray binaries
\citep[e.g.,][]{1982Natur.300..728A, 1991PhR...203....1B}.
They are key objects for studying the evolutionary history of close neutron star (NS) binary systems,
providing valuable insights into key mechanisms,
including the decay of NS surface magnetic field,
magnetic braking, the NS equation of state, and accretion efficiency
\citep[e.g.,][]{2001ApJ...557..958C, 2012MNRAS.425.1601T, 2019MNRAS.483.5595V, 2021ApJ...908..122G}.
Eclipsing MSPs are a class of close pulsar binaries with orbital periods of $\lesssim1.0$\,d,
in which the companion is ablated by the $\gamma$-ray irradiation and the energetic particles emitted by MSPs
\citep[e.g.,][]{1988Natur.334..225K, 1989ApJ...336..507R}.
Based on the properties of their companions,
eclipsing MSPs are typically divided into two subclasses:
black widows (BWs), with companion masses $\lesssim0.05\,M_\odot$,
and redbacks (RBs), with companion masses ranging from $\sim0.1-1.0\,M_\odot$
\citep[e.g.,][]{2013IAUS..291..127R, 2013ApJ...775...27C, 2014ApJ...786L...7B, 2015ApJ...814...74J, 2022MNRAS.515.2725G}.

With the increasing number of pulsar survey projects,
some peculiar eclipsing MSPs have been discovered.
For instance, BWs with planetary-mass companions have been identified
\citep[PSRs J$1719-1438$ and J$2322-2650$;][]{2011Sci...333.1717B, 2018MNRAS.475..469S},
which are thought to evolve into isolated MSPs within the Hubble timescale
\citep[e.g.,][]{2012A&A...541A..22V, 2022MNRAS.515.2725G}.
Additionally,
PSR J$1953+1844$ \citep[i.e., M71E;][]{cite-key},
an MSP binary with the shortest known orbital period was discovered by the
Five-hundred-meter Aperture Spherical radio Telescope (FAST)
during the FAST Galactic Plane Pulsar Snapshot survey
\citep[][]{2021RAA....21..107H}.
It has been suggested that this system originate from the evolution of ultracompact X-ray binaries (UCXBs) with He star donors
\citep[e.g.,][]{zonglin1, 2024MNRAS.527.7394G}.

Recently, FAST detected a new eclipsing MSP,
PSR J$1928+1815$, the first known pulsar with a He star companion
\citep[see][]{yang2025}.
This system features an orbital period of $0.15\,$d, companion mass of $\gtrsim1.0\,M_\odot$,
and a pulsar spin period of $10.55$\,ms.
The authors suggested that
this pulsar binary has just undergone common envelope (CE) ejection phase,
as indicated by its tight orbit,
and that the NS may be recycled to millisecond periods during the CE phase through hypercritical accretion.
However,
it remains still highly uncertain whether NSs can obtain the angular momentum
and spin up to millisecond periods during the CE process.
Alternatively,
PSR J$1928+1815$ is more likely formed after the Case BB mass transfer,
during which the NS can accrete material through the inner Lagrange point,
naturally leading to the formation of an MSP.
In this work,
we will systematically study the origin of eclipsing MSPs through He star donor channel.

Section \ref{sec:methods} describes on the numerical code
and the basic assumptions for the binary evolution simulations.
Section \ref{sec:results} presents the binary evolutionary
and binary population synthesis (BPS) results for
the formation of eclipsing MSPs with (evolved) He star companions.
In Section \ref{sec:discussion},
we give a relevant discussion,
followed by a summary in Section \ref{sec:summ}.

\section{Numerical methods and assumptions}\label{sec:methods}
We carried out detailed binary evolution calculations of NS + He star
systems by using the stellar evolution code Modules for Experiments
in Stellar Astrophysics
\citep[MESA, version 10398;][]{2018ApJS..234...34P}.
In our simulations,
the NSs are treated as point masses with initial masses of $1.4\,M_\odot$.
He star models are constructed assuming a typical Population I metallicity of $Z=0.02$,
with compositions of $98\%$ helium and $2\%$ metallicity.
To explore the systems that can produce eclipsing events,
we investigated the evolution of NS binaries with different initial companion masses
($M_{2}^{\rm i}=0.5-3.5\,M_\odot$),
and different initial orbital periods ($P_{\rm orb}^{\rm i}=0.04-2.00$\,d).
During the binary evolution,
the NS begins accreting material once the companion fills its Roche lobe,
and the mass-transfer rate is calculated using the scheme proposed by \citet{kolb1990A&A}.
In this work,
the fraction of mass lost from the vicinity of the NS is set to be $0.5$,
and we set the Eddington accretion rate $\dot M_{\rm Edd}=3\times10^{-8}\rm\,M_\odot\,\rm yr^{-1}$
\citep[e.g.,][]{2020ApJ...900L...8C, 2002MNRAS.331.1027D}.
Gravitational wave (GW) radiation has a significant impact on the orbital evolution of close binary systems,
and the orbital angular momentum loss due to the GW radiation
can be calculated by
\citep[see][]{1971ctf..book.....L}:
\begin{equation}
	\frac{d J_{\mathrm{GR}}}{d t}=-\frac{32 G^{7 / 2}}{5 c^5} \frac{M_{\mathrm{NS}}^2 M_{\mathrm{2}}^2\left(M_{\mathrm{NS}}+M_{\mathrm{2}}\right)^{1 / 2}}{a^{7 / 2}},
\end{equation}
in which the $G$, $c$ and $a$ are
the gravitational constant, the speed of light in vacuum
and the orbital separation, respectively.

Recent observations show that
the stripped stars have weak winds
\citep[$\lesssim10^{-9}\,M_\odot\,\rm yr^{-1}$;][]{2023Sci...382.1287D, 2023ApJ...959..125G}.
In this work,
we adopted the prescription of \citet[][]{2010MNRAS.404.1698J}
to calculate the stellar wind mass-loss rate ($\dot M_{\rm 2, stellar}$),
which is based on stellar wind analysis from sub-dwarf O stars (including He stars) to massive O stars:
\begin{equation}
\log \dot{M}_{\text {2, stellar}}=1.5 \log L / L_{\odot}-14.4,
\end{equation}
where $L$ is the surface luminosity.
This equation allows for relatively weaker stellar winds
\citep[see also][]{2010RAA....10..681W}.
We assumed that once the mass transfer ceases and the pulsar is spun up to a millisecond period,
the pulsar radiation sweeps across the companion star, thereby triggering the evaporation process.
We used the following formula
to calculate the mass-loss rate of companion caused by the evaporation
\citep[see][]{1992MNRAS.254P..19S}:
\begin{equation}
	\dot{M}_{\rm 2, evap}=-\frac{f}{2 \nu_{\rm2, esc }^2} L_{\mathrm{P}}\left(\frac{R_2}{a}\right)^2,
\end{equation}
where $f$, $\nu_{\rm2, esc }$ and $R_2$ are
the evaporation efficiency, the escape velocity at the donor surface
and the donor radius, respectively.
Previous studies suggested that the evaporation efficiency may be relatively low
\citep[e.g.,][]{2020MNRAS.495.3656G, 2022MNRAS.515.2725G},
therefore in this work we set $f = 0.01$.
$L_{\rm P} = 4\pi^2I\dot P_{\rm spin}/P_{\rm spin}^3$
is the spin-down luminosity of pulsars,
where $I=10^{45}\rm\,g\,cm^2$ is the pulsar moment of inertia,
$P_{\rm spin}$ and $\dot P_{\rm spin}$
are the pulsar spin period and its derivative, respectively.
We adopted initial values of $P_{\rm spin}=3$\,ms and $\dot P_{\rm spin}=1.0\times10^{-20}\rm\,s\,s^{-1}$,
and assumed a constant braking index of $n=3$
to calculate the evolution of $L_{\rm P}$
\citep[e.g.,][]{2013ApJ...775...27C, 2015ApJ...814...74J}.

We used the nuclear reaction network co\_burn in our simulations,
including helium, carbon and oxygen burning coupled by $57$ reactions.
The Type 2 Rosseland mean opacity tables provided by \citet[][]{1996ApJ...464..943I}
are adopted,
which is applicable to the enhanced carbon–oxygen caused by helium burning.
In our simulations,
we considered convective overshooting,
with the overshooting parameter set to $f_{\rm ov}=0.014$
\citep[e.g.,][]{2013ApJ...772..150J, Antoniadis2020A&A}.

\section{Formation of eclipsing MSPs with He-star companions} \label{sec:results}
\begin{figure*}
	\centering\includegraphics[width=0.6\textwidth]{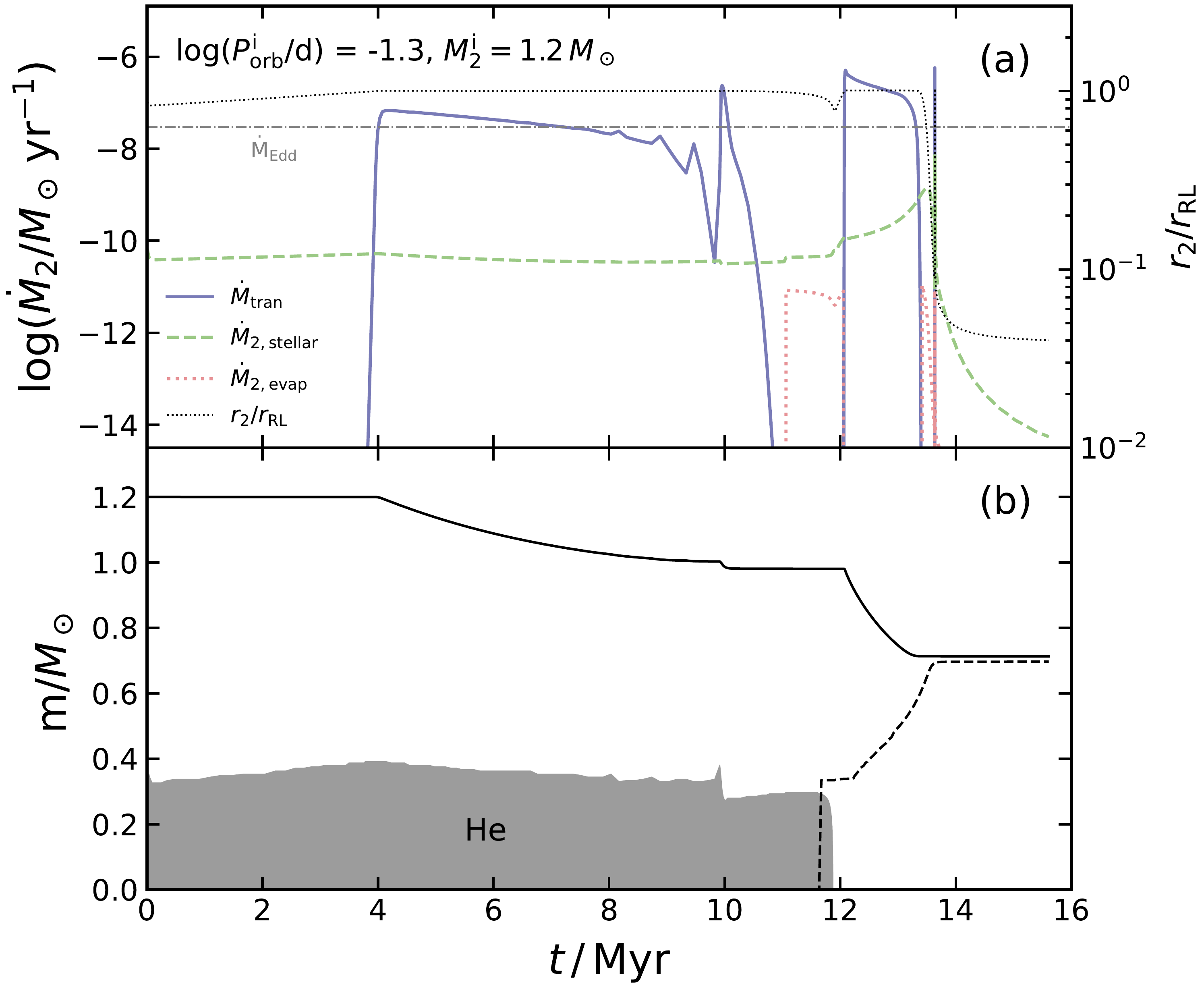}
	\caption{Case BA.
		The evolution of NS binary with
		log$(P_{\rm orb}^{\rm i}/\rm d) = -1.3$ and $M_2^{\rm i} = 1.2\,M_\odot$.
		Panel a: mass-transfer rate, stellar wind mass-loss rate and evaporation wind mass-loss rate as functions of stellar age.
		Panel b: Kippenhahn diagram of the $1.2\,M_\odot$ He star companion.
		The shaded area indicates convective regions induced by the He burning.
		The dashed line represents the CO core mass.}
	\label{fig:caseba}
\end{figure*}

\begin{figure*}
	\centering\includegraphics[width=0.6\textwidth]{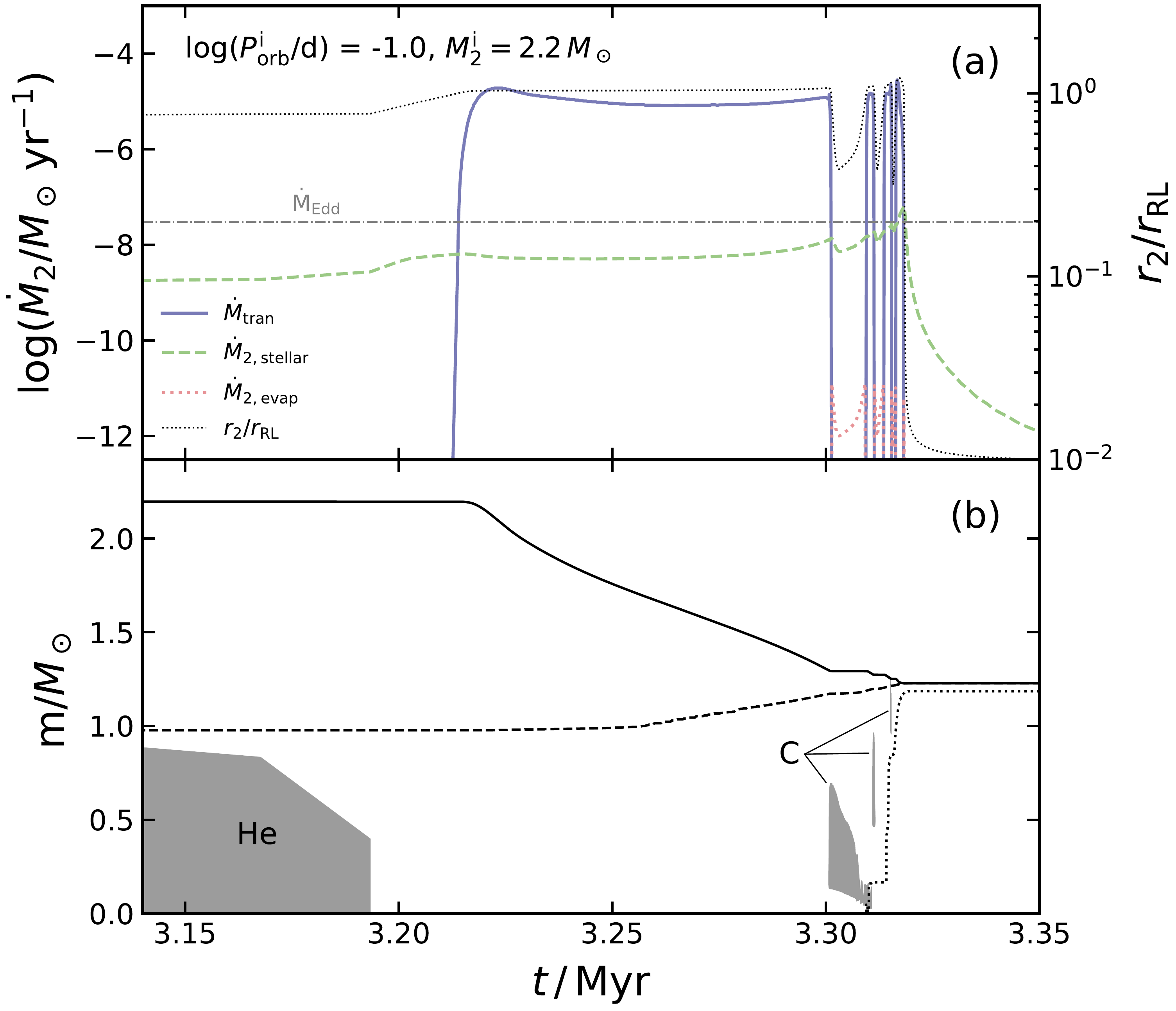}
	\caption{Case BB.
		Similar to Fig.\,\ref{fig:caseba}, but for NS binary with
		log$(P_{\rm orb}^{\rm i}/\rm d) = -1.0$ and $M_2^{\rm i} = 2.2\,M_\odot$.
		The shaded areas mark the convection regions caused by the He and C burning.
		The dashed and dotted lines represents the CO and ONe core masses, respectively.}
	\label{fig:casebb}
\end{figure*}

We performed a series of detailed binary evolution calculations of NS + He star
systems to investigate the formation of eclipsing MSPs.
Tables\,\ref{table:caseba}$-$\ref{table:casebb} summarize
the main evolutionary properties of some typical NS + He star binaries
that can produce the eclipsing MSPs,
in which the NS binaries undergo different mass-transfer processes,
i.e., Case BA and Case BB.
We present the calculated mass-transfer rates of NS binaries,
the Kippenhahn diagram and H-R diagram of companions in Figs\,\ref{fig:caseba}$-$\ref{fig:HR}.

\subsection{Case BA mass transfer} \label{sec:BA-mtran}
Case BA mass transfer is initiated in the NS + He star binary with
tight orbit $\lesssim0.06$\,d,
during which the He star companion is in the central helium-burning phase
\citep[e.g.,][]{2002MNRAS.331.1027D, 2009MNRAS.395..847W}.
Fig.\,\ref{fig:caseba}a shows the mass-transfer rate ($\dot M_{\rm tran}$)
as a function of stellar age ($t$) for an NS binary with
log($P_{\rm orb}^{\rm i}/\rm d)=-1.3$ and $M_{2}^{\rm i} = 1.2\,M_\odot$,
in which this system can evolve into an eclipsing MSP
after Case BA mass transfer.
At $t\sim3.80$\,Myr,
the He star companion fills its Roche lobe
as the orbit shrinks due to the GW radiation,
thereby leading to the first Roche-lobe overflow (RLOF) occurring on a nuclear timescale.
As the central He abundance decreases,
the companion contracts inside its Roche lobe, causing the binary to become detached.
At this moment,
the NS has accreted $\sim0.1\,M_\odot$ of material
and is fully recycled into an MSP.
In addition,
the orbital period and the companion mass are
$0.043$\,d and $0.98\,M_\odot$, respectively.
The thin black dotted line in Fig.\,\ref{fig:caseba}a
represents the evolution of the ratio of companion radius ($r_2$) to its Roche lobe radius ($r_{\rm RL}$).
During the detached phase,
the companion does not  contract significantly
and remains only slightly smaller than its Roche lobe, 
suggesting that pulsar radiation could evaporate material from the companion.
The evaporation process lasts for $\sim1.06$\,Myr.
Fig.\,\ref{fig:caseba}a shows the evolution of the calculated evaporation wind mass-loss rate
($\dot M_{\rm 2, evap}$) during the evaporation process.
$\dot M_{\rm 2, evap}$ is much lower than both
$\dot M_{\rm tran}$ and $\dot M_{\rm 2, stellar}$,
indicating that the evaporation wind has negligible influence on the evolutionary track of NS binaries.

\begin{figure*}
	\centering\includegraphics[width=0.6\textwidth]{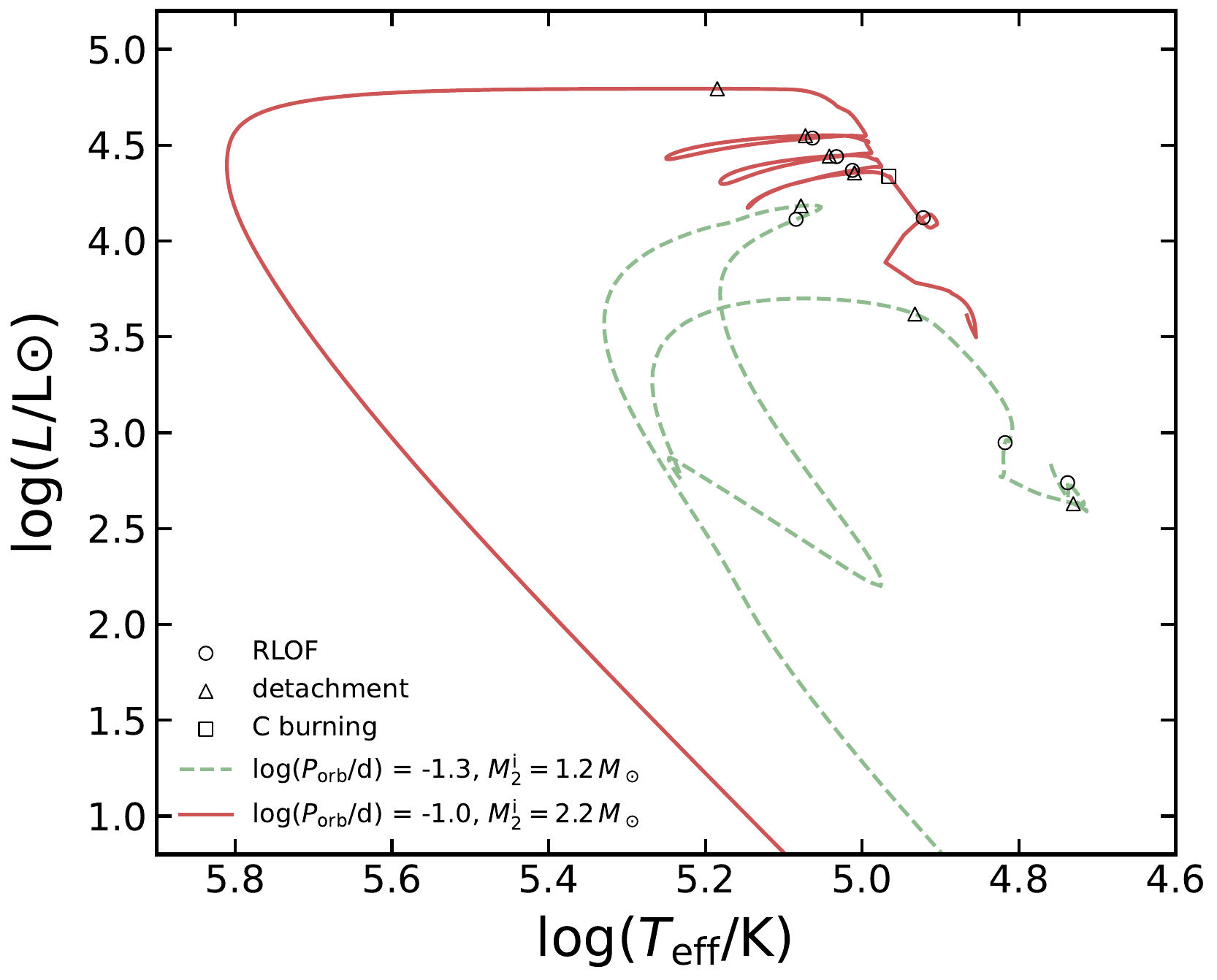}
	\caption{H-R diagram for the $1.2\,M_\odot$ He star companion in Case BA
		and the $2.2\,M_\odot$ He star companion in Case BB.
		The circles and triangles represent the onset
		of RLOF and detachment, respectively.
Square indicates the onset of C-shell flash.
}
	\label{fig:HR}
\end{figure*}

Fig.\,\ref{fig:caseba}b shows the evolution of convective region in the companion.
After central helium is exhausted,
the companion develops a $\sim0.34\,M_\odot$ CO core
surrounded by a $\sim0.64\,M_\odot$ He envelope.
The second RLOF is triggered at $t\sim12.1$\,Myr as the companion expands
due to the He-shell burning (so-called Case BAB),
during which the CO core mass gradually increases.
By $t\sim13.4$\,Myr,
the entire He envelope is almost removed and the NS binary becomes detached again.
At this stage,
the companion consists of a $\sim0.62\,M_\odot$ CO core
and a  $\sim0.09\,M_\odot$ He shell.
Subsequently,
the companion evolves toward the white dwarf (WD) cooling track.
In this case,
a thermal runaway is developed due to unstable He burning
on the companion surface during the cooling phase.
The green dashed line in Fig.\,\ref{fig:HR} represents the complete evolutionary track of
the $1.2\,M_\odot$ He star companion in the H-R diagram.
The He-shell flash triggers a dramatic and rapid luminosity increase,
after which the companion enters the cooling track again
and eventually evolves into a $\sim0.71\,M_\odot$ CO WD.

\subsection{Case BB mass-transfer} \label{sec:BB-mtran}
Fig.\,\ref{fig:casebb} shows the evolution of the mass-transfer rate (panel a)
and the convective structure (panel b) of the companion for an NS binary with
log($P_{\rm orb}^{\rm i}/\rm d)=-1.0$ and $M_2^{\rm i} = 2.2\,M_\odot$.
Following the central helium burning phase,
the He star companion develops a $\sim0.98\,M_\odot$ CO core
which is surrounded by a $\sim1.22\,M_\odot$ He envelope.
At $t\sim3.21$\,Myr,
the companion fills its Roche lobe due to the expansion caused by He-shell burning,
thereby triggering the Case BB RLOF.
From Fig.\,\ref{fig:casebb}a,
we can see that
Case BB RLOF lasts for $\sim0.09$\,Myr,
with the mass-transfer rates reaching highly super-Eddington values
($\sim10^3\dot M_{\rm Edd}$).
Previous studies suggest that Case BB RLOF allows NSs to accrete material
at mass-accretion rates of $\sim2-3$ times $\dot M_{\rm Edd}$
\citep[e.g.,][]{2014MNRAS.437.1485L, 2017ApJ...846..170T}.
\cite{2024MNRAS.530.4461G} also found that
NSs could accrete more material
if the residual rich-H envelope on the He star companion is considered.
Accordingly,
we assume that the NSs in Case BB RLOF could
increase their mass at accretion rates of $3\times\dot M_{\rm Edd}$.
In this example,
the NS accretes $\sim0.01\,M_\odot$ of material during Case BB,
sufficient to recycle it into an MSP.
During the detached phase,
the companion star maintains a relatively large radius
($r_2/r_{\rm RL}>0.4$) due to its thick helium envelope,
which allows evaporation events to occur.
Similar to Case BA,
our calculations of the evaporation wind (see Fig.\,\ref{fig:casebb}a)
show that its value is much lower than that of stellar wind,
indicating that the evaporation event has a negligible impact
on the binary evolutionary track and the final outcome.


The companion develops a $\sim1.17\,M_\odot$ CO core after Case BB RLOF,
exceeding the critical value for carbon ignition
\citep[$\sim1.05\,M_\odot$;][]{2015ApJ...807..184F}.
As the CO core contracts,
its temperature and central density progressively increase.
Fig.\,\ref{fig:casebb}b shows the
evolution of convective region for the companion.
The first C-shell flash takes place off-centre at mass coordinates of $\sim0.1\,M_\odot$,
due to the temperature inversion caused by the neutrino emission in the core center
\citep[e.g.,][]{2013ApJ...772..150J, 2015ApJ...807..184F}.
The red solid line in Fig.\,\ref{fig:HR} represents the complete evolutionary track of
the $2.2\,M_\odot$ He star companion in the H-R diagram.
Each episode of C-shell flash causes the companion to expand,
subsequently refilling its Roche lobe and triggering the next phase of mass transfer.
Following several C-shell flashes,
the companion transitions into a $\sim1.23\,M_\odot$ ONe core and enters to the WD cooling phase.

\subsection{Initial parameter of eclipsing MSPs with He star companions} \label{sec:space}
\begin{figure*}
	\centering\includegraphics[width=0.6\textwidth]{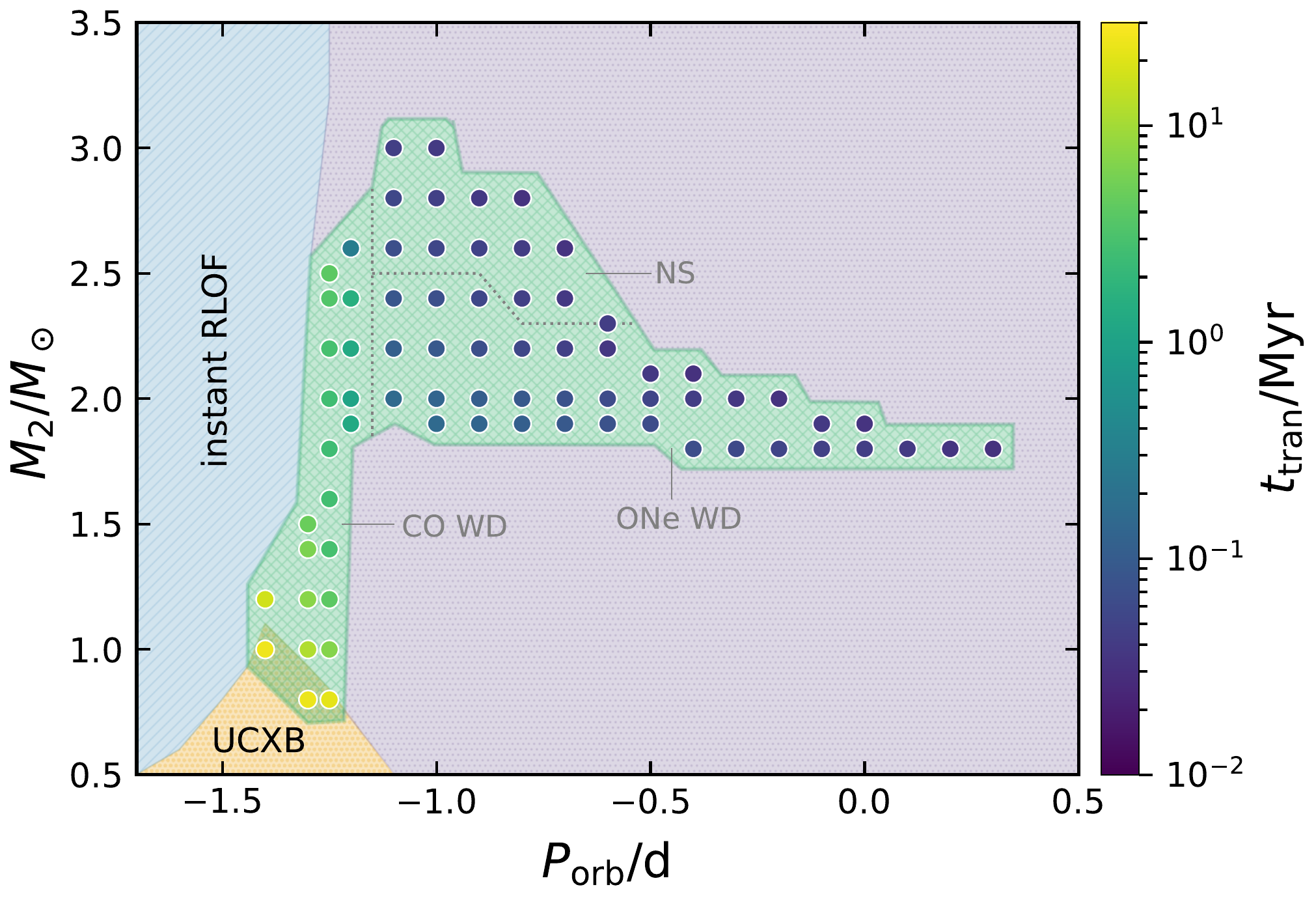}
	\caption{Initial parameter space (the green area) for producing eclipsing MSPs
		in the log$P_{\rm orb}^{\rm i}-M_2^{\rm i}$ plane.
		The solid circles in the green area represent the binaries
		that could evolve into eclipsing MSPs.
		The colorbar represents the duration of the first mass-transfer phase.
We also show the final evolutionary outcomes of the companions, i.e., CO WD, ONe WD and NS.
		The blue area indicates the NS binaries that RLOF begins when the He star is at the zero-age MS stage.
		The yellow area denotes the NS binaries that can form UCXBs.
		The purple area represents other NS binaries that can not form the eclipsing MSPs.}
	\label{fig:conf}
\end{figure*}

We calculated a large number of NS + He star systems for the formation of eclipsing MSPs,
thereby obtaining a dense model grid of binaries.
Tables\,\ref{table:caseba}$-$\ref{table:casebb} list
the main  evolutionary properties of NS+He star binaries that
can evolve into eclipsing MSPs.
Fig.\,\ref{fig:conf} shows the initial parameter space for producing
eclipsing MSPs in the log$P_{\rm orb}^{\rm i}-M_2^{\rm i}$ plane
(see the green area).
The eclipsing MSPs can form
if the initial parameters of NS + He star binaries fall within this contour.
The NS binaries with log($P_{\rm orb}^{\rm i}\rm /d)\lesssim-1.2$ will undergo Case BA RLOF,
which allows the NSs to accrete $\sim0.003-0.2\,M_\odot$ of material.
This implies that some systems undergoing Case BA may produce mildly recycled pulsars with
$10\lesssim P_{\rm spin}/\rm ms\lesssim 30$,
even though the majority of pulsars are expected to be fully recycled ($P_{\rm spin}\lesssim10\,$ms).
On the other hand,
if log($P_{\rm orb}^{\rm i}\rm /d)\gtrsim-1.2$,
the NS binaries will undergo Case BB RLOF,
during which the NSs typically accrete $\lesssim0.01\,M_\odot$ of material,
resulting in mildly recycled pulsars.
The final outcome of such system may be a pulsar + massive ONe WD binary.
Alternatively,
companions with relatively high initial masses
could undergo electron-capture supernovae (EC-SNe) or iron core-collapse supernovae (Fe CC-SNe),
leading to the formation of NSs.
If the system remains bound after the SN explosions,
it could eventually evolve into a double NS systems
\citep[e.g.,][]{2017ApJ...846..170T, 2018MNRAS.481.4009V, 2024MNRAS.530.4461G}.
The blue region on the left-hand boundary of the contour represents systems
in which RLOF begins while the He star is still at the zero-age main sequence (ZMS).
The yellow area corresponds to the NS binaries that can form UCXBs
\citep[e.g.,][]{2010NewAR..54...87N, 2021MNRAS.506.4654W, 2023MNRAS.521.6053L}.
The purple-shaded region represents other NS binaries that can not form eclipsing MSPs.
These systems either lack a detached phase or
fail to accrete sufficient material to spin up the NSs into MSPs.

\begin{figure*}
	\centering\includegraphics[width=0.6\textwidth]{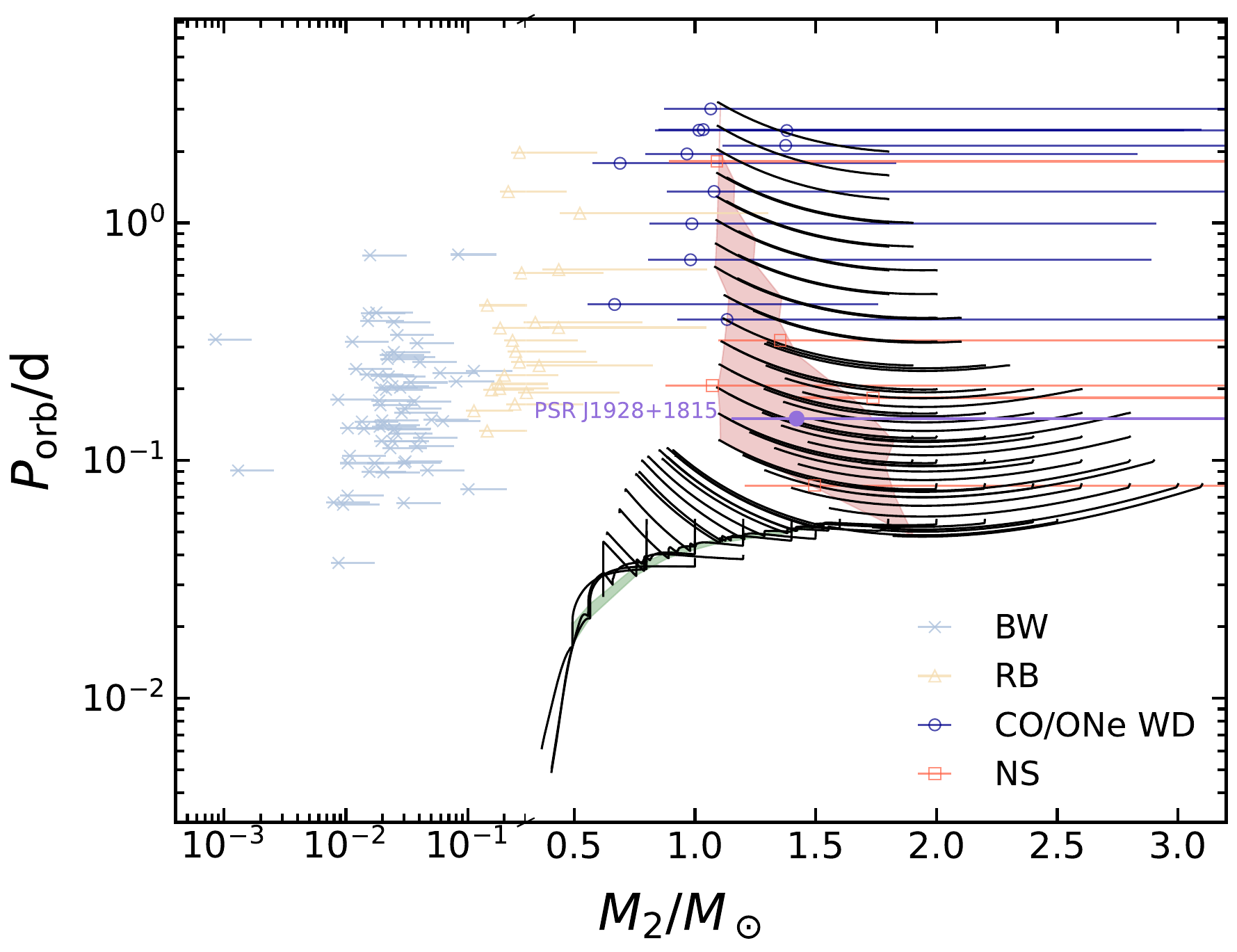}
	\caption{Evolutionary tracks for the binaries within the initial parameter space shown in Fig.\,\ref{fig:conf}.
	The green and red regions represent the eclipsing MSPs
	formed via Case BA and Case BB RLOF, respectively.
Circles and squares denote pulsar binaries containing CO/ONe WD and NS companions, respectively,
with orbital period $<4$\,d and spin period $\lesssim30$\,ms.
BW and RB samples are also included in the diagram.
Observational data are taken from the ATNF Pulsar Catalogue,
\href{http://www.atnf.csiro.au/research/pulsar/psrcat}{http://www.atnf.csiro.au/research/pulsar/psrcat}\citep[version 2.5.1, September 2024;][]{2005AJ....129.1993M}.
}
	\label{fig:mp}
\end{figure*}

Fig.\,\ref{fig:mp} shows the evolutionary tracks of NS + He star binaries
that can evolve into eclipsing MSPs in the $P_{\rm orb}-M_2$ diagram,
where the green and red regions represent
the eclipsing MSPs formed via Case BA and Case BB RLOF, respectively.
NS binaries evolving through Case BA typically form MSP + CO WD systems.
However,
the NS binaries with ultra-tight orbit and lower initial companion masses
may enter an UCXB phase
and eventually evolving into pulsar + planet-like systems
\citep[e.g.,][]{2010NewAR..54...87N, 2021MNRAS.506.4654W}.
For eclipsing MSPs formed via through Case BB,
the final outcomes depend on the metal core mass
formed within the companion.
These system may either leave behind ONe WDs
or undergo core collapse into NSs
via EC-SNe or Fe CC-SNe
\citep[e.g.,][]{2015MNRAS.451.2123T, 2024MNRAS.530.4461G, 2021ApJ...920L..36J}.

The purple symbol in Fig.\,\ref{fig:mp} marks the position of PSR J$1928+1815$ on the $P_{\rm orb}-M_2$ diagram.
Our simulations show that this system originates from Case BB channel,
with an initial orbital period of $\sim0.1\,$d.
Meanwhile,
its $10\,$ms spin period suggests that the NS needs to accrete $\sim0.01\,M_\odot$ of material,
which means that the initial mass of its He star companion is $\lesssim2.2\,M_\odot$ (see Table\,\ref{table:casebb}).
Accordingly,
we can further constrain the companion mass of the PSR J$1928+1815$ to be $\sim1.1-1.3\,M_\odot$.
This suggests that this system may eventually evolve into an MSP + massive ONe WD binary.
In Fig.\,\ref{fig:mp},
we show MSPs with orbital periods $<4$\,d that contain CO/ONe WD or NS companions.
Several of them fall within the red-shaded region,
indicating that they may have experienced a detached phase similar to that of PSR J$1928+1815$.
If the pulsar radiation can sweep across their companion during this phase,
the binaries may appear as eclipsing MSPs.
In addition,
we also predict another population of eclipsing pulsars with He star companions,
characterized by lower companion masses ($0.5\lesssim M_2/M_\odot \lesssim1.5$)
and tighter orbits ($0.01\lesssim P_{\rm orb}$/d $\lesssim0.05$).
These systems have undergone Case BA mass transfer,
during which the pulsars can be fully recycled or only mildly recycled.
This type of system has a relatively long lifetime ($\sim 1$\,Myr),
and we expect that future pulsar surveys will discover such pulsars.


\subsection{BPS results} \label{sec:bps}
\begin{figure*}
	\centering\includegraphics[width=0.6\textwidth]{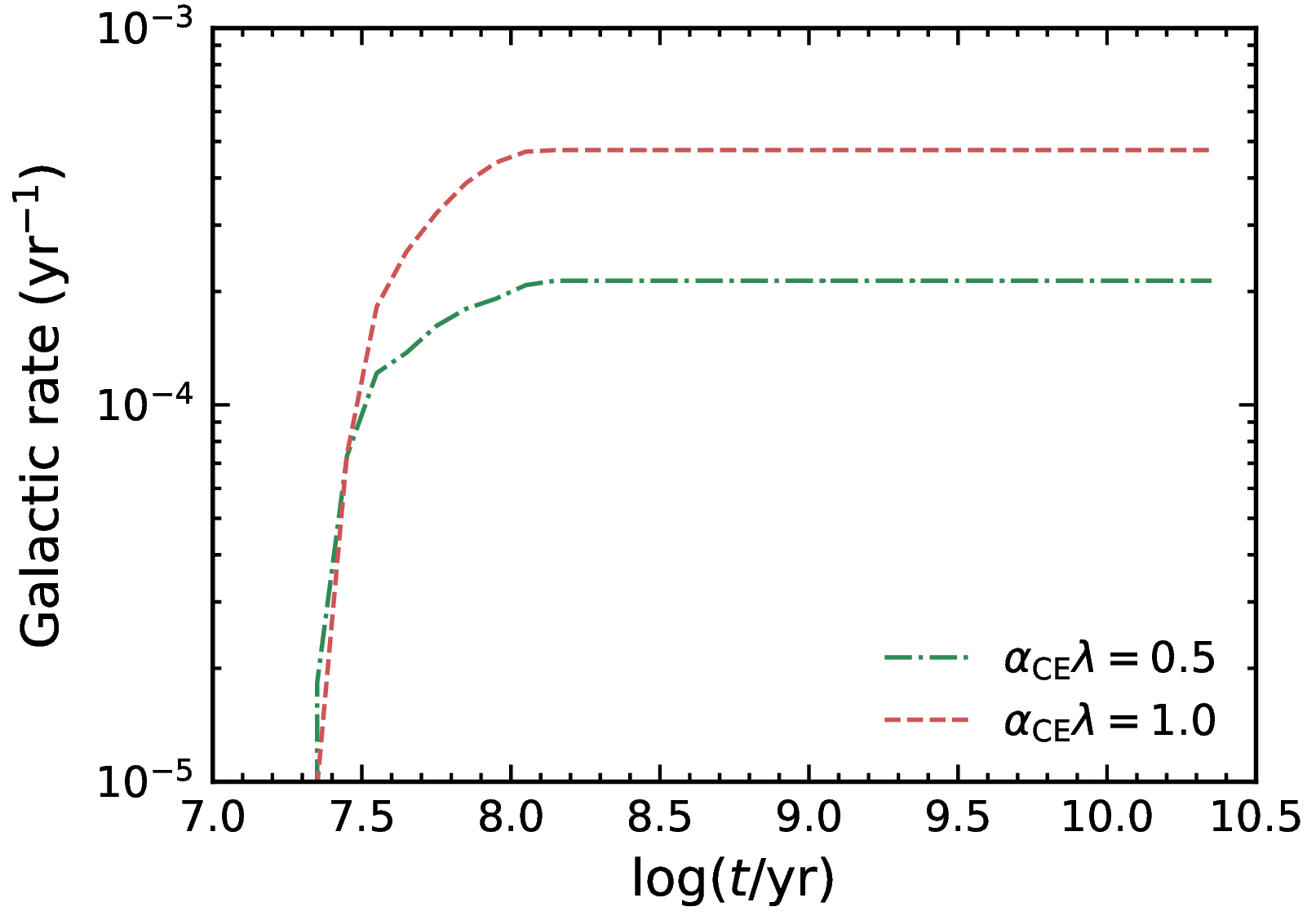}
	\caption{Evolution of the Galactic rates of eclipsing MSPs with He star companions
as a function of time,
assuming a constant Population I (Pop I) SFR of $5\,M_\odot\rm\,yr^{-1}$.
}
	\label{fig:rate}
\end{figure*}

Using the rapid binary star evolution (BSE) code
\citep[see][]{2002MNRAS.329..897H},
we carried out a series of Monte Carlo BPS calculations to
investigate the Galactic birthrate of eclipsing MSPs for the He star channel.
A sample of $1\times10^7$ primordial binaries are evolved until the formation of
NS + He star systems in each simulation.
We assumed that an eclipsing MSP
would be formed when the initial parameters of a NS+He star system
locate in the parameter space shown in Fig.\,\ref{fig:conf}.
The Monte Carlo BPS computations are based on the following initial parameters and fundamental assumptions
\citep[e.g.,][]{2009ApJ...701.1540W, 2018MNRAS.473.5352L, 2020RAA....20..161H}.

(1) We used the initial mass function provided by \cite{1979ApJS...41..513M}.

(2) A constant mass ratio distribution is adopted, expressed as n($q$) = 1, where $0 < q \leq 1$.

(3) All stars are considered to be members of binary systems with circular orbits.

(4) The initial distribution of orbital separations ($a$) is assumed to be constant in log\,$a$ for wide binaries,
and to decrease smoothly for close binaries.

(5) We assume a constant star formation rate of $5\,M_\odot$yr$^{-1}$ over the past 15 Gyr
\citep[e.g.,][]{1995MNRAS.272..800H, 2002MNRAS.329..897H}.

The NS+He star systems that produce eclipsing MSPs have tight orbits,
indicating that they most likely originate from the CE evolution.
The standard energy prescription is used to calculate the CE evolution
\citep[see][]{1984ApJ...277..355W},
where we combine the CE ejection efficiency ($\alpha_{\rm CE}$) and the stellar structure parameter ($\lambda$)
into a free parameter ($\alpha_{\rm CE}\lambda$), which is set to $0.5$ and $1.0$ in our simulations
\citep[e.g.,][]{2010MNRAS.401.2729W, 2018MNRAS.473.5352L, 2024MNRAS.534.3400L}.

Fig.\,\ref{fig:rate} shows the birthrate evolution of the eclipsing MSPs from He star donor channel
with different values of $\alpha_{\rm CE}\lambda$,
The BPS calculations estimate the theoretical eclipsing MSP rate in the Galaxy
to be $\sim2.1-4.7\times10^{-4}\rm\,yr^{-1}$.
With the evaporation timescales of $\sim0.6$\,Myr (Case BA) and $\sim0.01$\,Myr (Case BB),
we estimate that there could be $\sim55-150$ eclipsing MSPs with He star in the Galaxy.
Moreover,
pulsars like PSR J$1928+1815$ that have undergone Case BB mass transfer have a short evaporation duration,
leading to that their number in the Galaxy is likely to be $\lesssim10$.
This naturally accounts for their scarce detection.

\section{Discussion}\label{sec:discussion}

It is generally believed that
the accretion torque during the mass transfer can efficiently align the spin axis of MSPs
with the orbital angular momentum vector within $\lesssim 0.01$\,Myr
\citep[e.g.,][]{1991PhR...203....1B, 2021MNRAS.505.1775B}.
However,
\citet{yang2025} suggested that PSR J$1928+1815$ has a significant misalignment angle,
although the authors have not derived its value.
\citet{2023MNRAS.519.2951T}
proposed that the misalignment angle in MSP binaries may be explained by the thermonuclear rocket model proposed by \citet{2021ApJ...909..161H}.
In this scenario,
the misalignment angle may be produced by
the multiple weak kicks resulting from H-shell flashes occurring on the surface of the He WD companion.
Our simulations indicate that PSR J$1928+1815$ originates from the Case BB,
where the companion undergoes multiple C-shell flashes.
These flashes may generate multiple kicks,
potentially leading to the observed misalignment angle.
The kick velocity can be calculated by \citep[see][]{2021ApJ...909..161H}:
\begin{equation}
V_{\mathrm{k}} \simeq  8 \mathrm{~km} \mathrm{~s}^{-1}\left(\frac{\Delta M}{10^{-3} M_{\odot}}\right)\left(\frac{V_{\mathrm{ej}}}{2.5 \times 10^3 \mathrm{~km} \mathrm{~s}^{-1}}\right) 
 \left(\frac{0.3 M_{\odot}}{M_{\mathrm{2}}}\right),
\end{equation}
in which $\Delta M$, $V_{\rm ej}$ and $M_2$ are the
ejected mass during shell flashes, the ejecta velocity and the companion mass, respectively.

To obtain $\Delta M$, $V_{\rm ej}$ and $M_2$,
we simulated the evolution of a NS + $2.2\,M_\odot$ He star system with an initial $P_{\rm orb}$ of $0.1\,$d.
As shown in Fig.\,\ref{fig:casebb}b,
the C-shell flashes could occur near the surface of the companion.
Thus,
we adopted the super-Eddington wind during the C-shell flashes,
which is appropriate for the shell flash similar to nova outburst
\citep[e.g.,][]{2017A&A...604A..31W, 2022A&A...660A..53G}.
Our simulation results show that
C-shell flashes can last for $\lesssim10000\,$yr,
during which the companion mass, the mass loss, and the wind velocity are $\sim1.2\,M_\odot$,
$\sim0.1\,M_\odot$, and $\sim25000\,$km/s, respectively.
Since the kicks added to the companion are instantaneous in a realistic scenario,
we assume that the companion undergoes $10^{5}$ outbursts during the C-shell flash phase,
each with a short duration of $0.1$\,yr, a mass loss of \( 10^{-6}\,M_{\odot} \),
and a randomly oriented kick of $0.02$\,km/s.

We calculated the misalignment angle after the multiple outbursts
where the basic method is similar to that of \citet{2023MNRAS.519.2951T}.
The results show that
the maximum misalignment angle is only $\sim1^{\circ}$.
Meanwhile,
a larger misalignment angle is preferentially associated with a higher eccentricity ($e\lesssim0.01$),
whereas \citet{yang2025} suggested that PSR J$1928+1815$ has a circular orbit ($e < 3 \times 10^{-5}$).
In the more realistic scenario,
the instantaneous kick may be triggered on a shorter timescale.
It is reasonable to assume that
the each outburst timescale is much shorter than the orbital period of PSR J$1928+1815$ (i.e., $0.15$\,d),
which would lead to a lower misalignment angle.
We therefore conclude that
the multiple kicks induced by C-shell flashes may have negligible effects on the misalignment angle and the orbit.

We note that during the C-shell flashes,
the companion undergoes strong winds with a mass-loss rate of
$\sim10^{-5} M_{\odot}\rm\,yr^{-1}$ under the super-Eddington wind scheme,
while mass transfer ceases.
This suggests that the NS can accrete material from its companion through wind accretion
\citep[e.g.,][]{1981A&A...102...36W, 1988ApJ...335..862F, 2015ApJ...803...41M}.
During the wind accretion,
the direction of angular momentum captured by NS is less stable compared to the case of accretion through the inner Lagrange point.
If the spin-up torque changes randomly over time,
the spin axis of NS may undergo a non-linear, random-walk-like evolution
\citep[see][]{2021MNRAS.505.1775B}.
The key advantage of this model is that
it may produce a high misalignment angle
without inducing significant orbital eccentricity,
which is consistent with the observed circular orbit.
However,
it remains unclear whether the misalignment angle in the wind accretion model
undergoes significant changes during the C-shell flash phase,
which occurs on a timescale of $\lesssim0.01$\,Myr.
If this mechanism is at work,
eclipsing MSPs formed via Case BB may generally exhibit a noticeable misalignment angle. 
In contrast,
since the companion in Case BA do not experience strong stellar winds,
eclipsing MSPs formed via this pathway are not expected to exhibit significant misalignment angles.


\section{Summary}\label{sec:summ}
PSR J$1928+1815$ is the first discovered eclipsing MSP with a He star companion.
We investigated the formation of the newly recognized subclass of
eclipsing MSPs with (evolved) He star companions
through the He star donor channel,
and identified two evolutionary pathways:

(1) Case BA:
In close-orbit ($0.04\lesssim P_{\rm orb}/\rm\,d\lesssim 0.06$) NS + He star companion binaries that undergo Case BA and Case BAB mass transfer,
the evaporation process may take place during the detached phase between these two episodes.
Eclipsing MSPs formed through Case BA typically have companions with masses of $0.5-1.5\,M_{\odot}$
and short orbital periods of $0.01-0.05$\,d,
where the pulsar may be either fully recycled or mildly recycled.

(2) Case BB:
For relatively wider orbit ($0.08\lesssim P_{\rm orb}/\rm\,d\lesssim 2.0$) NS + He star binaries,
if the companion star develops a sufficiently massive CO core after experiencing Case BB,
it could trigger C-shell flashes, leading to subsequent multiple mass-transfer phases.
The radio eclipses may arise during the detached phase between these mass-transfer episodes.
The MSPs formed through Case BB typically exhibit companions with masses of $1.0-2.0\,M_{\odot}$
and orbital periods ranging from $0.05-2.0$\,d, where the pulsar is generally mildly recycled.

We conclude that PSR J$1928+1815$ originated from an NS + He star donor binary
that can undergo Case BB mass transfer,
with an initial orbital period of $\sim0.1\,$d.
In addition,
we obtained initial parameter spaces for producing the eclipsing MSPs with He star companions,
and estimated their birth rates ($\sim2.1-4.7\times10^{-4}\rm\,yr^{-1}$) and the numbers ($55-150$)
in the Galaxy using the BPS approach.
We also explored the properties of C-shell flashes in the Case BB channel,
and found that the instantaneous kicks induced by C-shell flashes are likely negligible for the binary system.
On the other hand,
the C-shell flashes could drive strong stellar winds,
potentially resulting in a significant misalignment angle of the pulsar through wind accretion.


\begin{acknowledgments}
This study is supported by the National Natural Science Foundation of China (Nos 12041301, 12090040/12090043, 12121003, 12225304, 12273105, 12288102 and 12403035), the National Key R\&D Program of China (Nos 2021YFA0718500, 2021YFA1600404, 2021YFA1600403 and 2021YFA1600400),
the Western Light Project of CAS (No. XBZG-ZDSYS-202117), the Yunnan Revitalization Talent Support Program (Yunling Scholar Project), the Yunnan Fundamental Research Project (No 202501AS070005), the International Centre of Supernovae, Yunnan Key Laboratory (No. 202302AN360001),
the Youth Innovation Promotion Association CAS (No. 2021058), the Yunnan Revitalization Talent Support Program (Young Talent project),
the Yunnan Fundamental Research Projects (Nos 202401AV070006, 202201AW070011 and 202201BC070003),
the Postdoctoral Fellowship Program of CPSF under Grant Number GZB20240307,
the China Postdoctoral Science Foundation under Grant Number 2024M751375 and 2024T170393,
and the Jiangsu Funding Programme for Excellent Postdoctoral Talent under Grant Number 2024ZB705.
\end{acknowledgments}

\bibliography{1bib}{}
\bibliographystyle{aasjournal}

\appendix
\section{Evolutionary properties of NS + He star companions}

\begin{table*}
	\centering

	\caption{
		Evolutionary properties of the NS+He star binaries with  different initial secondary masses and initial orbital periods.
The columns (from left to right): the initial orbital period and the initial mass of the He star companion;
the amount of mass transferred from the He star, the duration and
the maximum mass-transfer rate during the Case BA mass transfer;
the duration of the detached phase;
the amount of mass transferred from the He star, the duration and
the maximum mass-transfer rate during the case BAB mass transfer;
the final orbital period and companion mass;
and the final fate of companion star.
	}
	\label{table:caseba}
\footnotesize
	\begin{tabular}{cccccccccccccccc}
		\toprule
		$P_{\rm orb}^{\rm i}$ &$M_{\rm He}^{\rm i}$  & Case BA& $\Delta t_{\rm Case\,BA}$
		&$\dot M_{\rm tran, BA}$&$\Delta t_{\rm detached}$&Case BAB&$\Delta t_{\rm Case\,BAB}$&$\dot M_{\rm tran, BAB}$&$P_{\rm orb}$&$M_{\rm He}^{\rm f}$&Final fate&$M_{\rm NS, f}$\\
		
		(d)&($M_\odot$)&$\Delta M/M_\odot$&(Myr)&($M_\odot\,$yr$^{-1}$)&(Myr)&$\Delta M/M_\odot$&(Myr)&($M_\odot\,$yr$^{-1}$)&(d)&($M_\odot)$&&($M_\odot)$\\
		
		\hline
	$0.040$&$1.0$&$0.22$&$24.65$&$4.79\times10^{-8}$&$1.05$&$>6.98$e-2&$>4.97$&$2.66\times10^{-7}$&$-
	$&$\textless0.41$&CO WD&$>1.69$\\
	
		   &$1.2$&$0.21$&$16.75$&$1.50\times10^{-7}$&$1.42$&$5.27$e-2&$2.66$&$1.31\times10^{-7}$&$0.045$&$0.62$&CO WD&$1.66$\\
		   
		\hline
		
	$0.050$&$0.8$&$0.15$&$24.09$&$2.44\times10^{-8}$&$0.96$&$>5.36$e-2&$>2.65$&$1.46\times10^{-7}$&$-
	$&$0.37$&CO WD&$>1.61$\\
	
		  &$1.0$&$0.10$&$11.49$&$1.59\times10^{-7}$&$1.44$&$5.27$e-2&$2.33$&$1.68\times10^{-7}$&$0.05$&$0.64$&CO WD&$1.55$\\
		  
		 &$1.2$&$0.10$&$7.34$&$2.38\times10^{-7}$&$1.06$&$3.72$e-2&$1.36$&$5.06\times10^{-7}$&$0.076$&$0.71$&CO WD&$1.54$\\
		 
		&$1.4$&$0.10$&$6.18$&$7.02\times10^{-7}$&$0.91$&$2.33$e-2&$0.83$&$7.89\times10^{-7}$&$0.09$&$0.77$&CO WD&$1.53$\\
		
		&$1.5$&$0.10$&$4.88$&$1.49\times10^{-7}$&$0.80$&$2.46$e-2&$0.86$&$1.61\times10^{-6}$&$0.10$&$0.78$&CO WD&$1.53$\\
		
		\hline
		
 $0.056$&$0.8$& $0.11$& $21.34$& $2.60\times10^{-7}$& $0.73$& $>6.64$e-2& $>5.34$& $2.13\times10^{-7}$& $-$& $0.41$& CO WD& $>1.57$\\
		
		&$1.0$& $4.32$e-2& $6.81$& $3.70\times10^{-7}$& $1.30$& $3.86$e-2& $1.52$& $2.96\times10^{-7}$& $0.062$& $0.69$& CO WD& $1.48$\\
		
		&$1.2$& $4.02$e-2& $3.97$& $3.90\times10^{-7}$& $0.85$& $2.75$e-2& $0.98$& $8.86\times10^{-7}$& $0.088$& $0.76$& CO WD& $1.47$\\
		
		&$1.4$& $4.61$e-2& $2.85$& $9.53\times10^{-8}$& $0.72$& $2.08$e-2& $0.73$& $2.19\times10^{-6}$& $0.10$& $0.81$& CO WD& $1.47$\\
		
		&$1.6$& $4.90$e-2& $2.69$& $1.77\times10^{-7}$& $0.50$& $1.57$e-2& $0.54$& $3.27\times10^{-6}$& $0.11$& $0.85$& CO WD& $1.46$\\
		
		&$1.8$& $4.99$e-2& $2.46$& $3.53\times10^{-7}$& $0.52$& $1.31$e-2& $0.45$& $3.83\times10^{-6}$& $0.11$& $0.89$& CO WD& $1.46$\\
		
		&$2.0$& $5.46$e-2& $2.57$& $8.65\times10^{-7}$& $0.49$& $1.23$e-2& $0.42$& $3.73\times10^{-6}$& $0.11$& $0.91$& CO WD& $1.47$\\
		
		&$2.2$& $6.28$e-2& $2.84$& $3.41\times10^{-6}$& $0.51$& $1.36$e-2& $0.47$& $3.52\times10^{-6}$& $0.11$& $0.92$& CO WD& $1.48$\\
		
		&$2.4$& $7.35$e-2& $3.41$& $1.72\times10^{-5}$& $0.56$& $1.64$e-2& $0.56$& $2.91\times10^{-6}$& $0.10$& $0.89$& CO WD& $1.49$\\
		
		&$2.5$& $5.44$e-2& $3.94$& $3.26\times10^{-5}$& $0.58$& $1.80$e-2& $0.62$& $2.73\times10^{-6}$& $0.10$& $0.87$& CO WD& $1.47$\\
		
		\hline
		
$0.063$&$1.9$&$6.38$e-3&$1.25$&$4.89\times10^{-7}$&$0.34$&$6.63$e-3&$0.23$&$1.00\times10^{-5}$&$0.101$&$1.02$&CO WD&$1.41$\\

&$2.0$&$1.00$e-2&$1.08$&$3.77\times10^{-7}$&$0.34$&$6.70$e-3&$0.23$&$1.14\times10^{-5}$&$0.102$&$1.00$&CO WD&$1.42$\\

&$2.2$&$2.06$e-2&$1.25$&$3.16\times10^{-6}$&$0.38$&$6.85$e-3&$0.23$&$6.19\times10^{-6}$&$0.103$&$1.03$&CO WD&$1.43$\\

&$2.4$&$2.87$e-2&$1.57$&$1.66\times10^{-5}$&$0.43$&$9.47$e-3&$0.32$&$4.86\times10^{-6}$&$0.0938$&$1.04$&CO WD&$1.44$\\

&$2.6$&$3.19$e-3&$0.29$&$5.17\times10^{-5}$&$2.54$&$1.33$e-2&$0.46$&$2.11\times10^{-6}$&$0.0933$&$0.982$&CO WD&$1.42$\\
		
		\hline
	\end{tabular}
\end{table*}
\newpage

\small
\renewcommand{\arraystretch}{0.7}

\begin{longtable}{cccccccccc}
    \caption{
Similar to Table\,\ref{table:caseba}, but for Case BB mass transfer.}
    \label{table:casebb}\\
    
    \hline
    \hline
    $P_{\rm orb}^{\rm i}$ &$M_{\rm He}^{\rm i}$  & Case BB& $\Delta t_{\rm Case\,BB}$
    &$\dot M_{\rm tran, BB}$&$\Delta t_{\rm detached}$&$P_{\rm orb}$&$M_{\rm He}$&Final fate&$M_{\rm NS}$\\
    
    (d)&($M_\odot$)&$\Delta M/M_\odot$&(yr)&($M_\odot\,$yr$^{-1}$)&(yr)&(d)&($M_\odot$)& &($M_\odot$)\\
    \hline
    \endfirsthead
    
    \multicolumn{10}{l}{\textit{Continued from previous page}} \\
    \hline
    $P_{\rm orb}^{\rm i}$ &$M_{\rm He}^{\rm i}$  & Case BB& $\Delta t_{\rm Case\,BB}$
    &$\dot M_{\rm tran, BB}$&$\Delta t_{\rm detached}$&$P_{\rm orb}$&$M_{\rm He}$&Final fate&$M_{\rm NS}$\\
    
    (d)&($M_\odot$)&$\Delta M/M_\odot$&(yr)&($M_\odot\,$yr$^{-1}$)&(yr)&(d)&($M_\odot$)& &($M_\odot$)\\
    \hline
    \endhead
    
    \hline
    \endfoot
    
    \hline
    \endlastfoot
	
		$0.079$& $2.0$& $1.37$e-2& $1.55$e5& $1.26\times10^{-5}$& $1.76$e4& $0.121$& $1.106$&&$1.414$\\
&   *  & $8.17$e-5& $1.32$e3& $5.45\times10^{-6}$& $4.46$e3&
$0.121$& $1.104$&& $1.414$\\
&   ** & $1.68$e-4& $2.60$e3& $2.80\times10^{-6}$& -& $0.122$& $1.101$&& $1.414$\\

&$2.2$& $9.62$e-3& $1.10$e5& $2.04\times10^{-5}$& $1.08$e4& $0.100$& $1.245$&-&$1.410$\\
&   *  & $1.11$e-4& $1.59$e3& $1.13\times10^{-5}$& $3.22$e3& $0.101$& $1.236$&-& $1.410$\\
&   ** & $1.92$e-4& $2.49$e3& $1.88\times10^{-5}$& $1.23$e3& $0.103$& $1.217$&-& $1.410$\\
&   ***& $1.78$e-4& $2.37$e3& $1.76\times10^{-5}$&        -& $0.105$& $1.201$&ONe WD& $1.410$\\

&$2.4$& $7.45$e-3& $8.52$e4& $3.18\times10^{-5}$& $1.17$e4& $0.085$& $1.378$&-& $1.407$\\
&   * & $1.45$e-4& $1.97$e3& $2.60\times10^{-5}$& $2.43$e3& $0.087$& $1.349$&-& $1.408$\\
&   **& $1.26$e-4& $1.72$e3& $4.99\times10^{-5}$& $1.82$e3& $0.088$& $1.325$&-& $1.408$\\
&  ***& $7.83$e-5& $9.55$e2& $1.03\times10^{-4}$& -       & $0.091$& $1.290$&ONe WD&$1.408$\\

&$2.6$& $5.88$e-3& $6.76$e4& $4.83\times10^{-5}$& $9.40$e3& $0.071$& $1.521$&-&$1.406$\\
&   * & $1.81$e-4& $2.88$e3& $4.16\times10^{-5}$& $1.72$e3& $0.073$& $1.472$&-& $1.406$\\
&   **& $8.90$e-5& $1.13$e3& $1.06\times10^{-2}$& -       & $0.077$& $1.402$&EC-SN &$1.406$\\

&$2.8$& $4.68$e-3& $5.43$e4& $7.23\times10^{-5}$& $5.40$e3& $0.060$& $1.67$&-& $1.405$\\
&   * & $2.21$e-4& $3.95$e3& $6.52\times10^{-5}$& $1.00$e3& $0.062$& $1.607$&-& $1.405$\\
&   **& $2.55$e-5& $3.93$e2& $3.75\times10^{-2}$& -       & $0.063$& $1.559$&Fe CC-SN& $1.405$\\

&$3.0$& $3.80$e-3& $4.44$e4& $1.07\times10^{-4}$& $1.96$e3& $0.052$& $1.824$&-& $1.404$\\
&   * & $2.57$e-4& $5.06$e3& $8.62\times10^{-5}$& $1.49$e2& $0.052$& $1.751$&-& $1.404$\\
&   **& $1.12$e-5& $2.62$e2& $5.70\times10^{-3}$& -       & $0.052$& $1.737$& Fe CC-SN& $1.404$\\[2pt]

\hline\\[-8pt]

$0.1$& $1.9$& $1.22$e-2& $1.52$e5& $9.01\times10^{-6}$& $1.75$e4& $0.156$& $1.104$&-& $1.412$\\
&   *  & $9.05$e-5& $1.36$e3& $5.11\times10^{-6}$& $4.27$e3& $0.157$& $1.101$&-& $1.412$\\
&   ** & $1.43$e-4& $2.27$e3& $2.63\times10^{-6}$&-        & $0.157$& $1.099$& ONe WD& $1.412$\\

& $2.0$& $1.14$e-2& $1.31$e5& $1.16\times10^{-5}$& $1.24$e4& $0.150$& $1.137$&-& $1.411$\\
&   *  & $8.31$e-5& $1.23$e3& $7.07\times10^{-6}$& $3.90$e3& $0.151$& $1.133$&-& $1.411$\\
&   ** & $2.68$e-4& $3.54$e3& $5.61\times10^{-6}$&        -& $0.152$& $1.124$& ONe WD &$1.412$\\
& $2.2$& $7.85$e-3& $9.07$e4& $1.90\times10^{-5}$& $7.81$e3& $0.123$& $1.294$&-& $1.408$\\
&    * & $1.61$e-4& $2.18$e3& $1.48\times10^{-5}$& $2.05$e3& $0.126$& $1.274$&-& $1.408$\\
&    **& $1.56$e-4& $1.99$e3& $2.33\times10^{-5}$& $9.03$e2& $0.128$& $1.251$&-& $1.408$\\
&   ***& $1.62$e-4& $1.94$e3& $2.83\times10^{-5}$& -       & $0.131$& $1.229$& ONe WD& $1.408$\\

& $2.4$& $5.91$e-3& $7.08$e4& $3.00\times10^{-5}$& $9.06$e3& $0.103$& $1.450$&-& $1.406$\\
&    * & $1.64$e-4& $2.32$e3& $3.52\times10^{-5}$& $1.70$e3& $0.106$& $1.409$&-& $1.406$\\
&    **& $1.76$e-4& $2.13$e3& $2.07\times10^{-4}$& -       & $0.113$& $1.332$& ONe WD& $1.406$\\

& $2.6$& $4.66$e-3& $5.47$e4& $4.57\times10^{-5}$& $5.56$e3& $0.088$& $1.596$&-& $1.405$\\
&    * & $2.06$e-4& $3.20$e3& $5.36\times10^{-5}$& $9.88$e2& $0.091$& $1.536$&-& $1.405$\\
&    **& $6.39$e-5& $8.19$e2& $8.37\times10^{-3}$& -       & $0.096$& $1.430$& EC-SN& $1.405$\\

& $2.8$& $3.70$e-3& $4.61$e4& $6.48\times10^{-5}$& $2.77$e3& $0.076$& $1.753$&-& $1.404$\\
&    * & $2.48$e-4& $4.64$e3& $2.59\times10^{-3}$& -       & $0.078$& $1.635$& Fe CC-SN& $1.404$\\

& $2.9$& $3.30$e-3& $3.94$e4& $7.88\times10^{-5}$& $1.33$e3& $0.070$& $1.829$&-& $1.403$\\
&    * & $2.47$e-4& $3.99$e3& $1.32\times10^{-2}$& -       & $0.072$&
$1.714$& Fe CC-SN & $1.404$\\

\hline\\[-8pt]

$0.126$& $1.9$& $1.14$e-2& $1.32$e5& $9.01\times10^{-6}$& $1.43$e4& $0.202$& $1.095$&-& $1.411$\\
&    * & $2.21$e-4& $3.19$e3& $2.42\times10^{-6}$& -       & $0.203$& $1.092$& ONe WD& $1.412$\\

& $2.0$& $9.23$e-3& $1.07$e5& $1.25\times10^{-5}$& $1.26$e4& $0.184$& $1.166$&-& $1.409$\\
&    * & $8.33$e-5& $1.20$e3& $8.13\times10^{-6}$& $4.36$e3& $0.186$& $1.161$&-& $1.409$\\
&    **& $2.58$e-4& $3.08$e3& $7.60\times10^{-6}$& $1.54$e3& $0.189$& $1.145$&-& $1.410$\\
&   ***& $8.36$e-5& $1.23$e3& $4.04\times10^{-6}$& -       & $0.190$& $1.143$& ONe WD& $1.410$\\

& $2.2$& $5.60$e-3& $6.77$e4& $1.94\times10^{-5}$& $9.00$e3& $0.146$& $1.381$&-& $1.406$\\
&    * & $1.67$e-4& $2.18$e3& $2.70\times10^{-5}$& $1.90$e3& $0.150$& $1.345$&-& $1.406$\\
&    **& $1.23$e-4& $1.61$e3& $5.11\times10^{-5}$& $2.54$e2& $0.153$& $1.319$&-& $1.406$\\
&   ***& $9.07$e-5& $1.08$e3& $9.03\times10^{-5}$& -       & $0.159$& $1.281$&ONe WD& $1.406$\\

& $2.4$& $4.63$e-3& $5.52$e4& $2.98\times10^{-5}$& $6.41$e3& $0.127$& $1.509$&-& $1.405$\\
&    * & $1.83$e-4& $2.74$e3& $4.31\times10^{-5}$& $1.16$e3& $0.131$& $1.456$&-& $1.405$\\
&    **& $1.35$e-4& $1.64$e3& $5.73\times10^{-4}$& -       & $0.140$& $1.360$&ONe WD& $1.405$\\

& $2.6$& $4.08$e-3& $4.87$e4& $4.73\times10^{-5}$& $5.30$e3& $0.112$& $1.610$&-& $1.404$\\
&    * & $2.33$e-4& $3.82$e3& $5.84\times10^{-5}$& $1.04$e3& $0.115$& $1.543$&-& $1.404$\\
&    **& $6.27$e-5& $7.99$e2& $2.45\times10^{-3}$& -       & $0.120$& $1.471$&Fe CC-SN& $1.404$\\

& $2.8$& $3.10$e-3& $3.77$e4& $6.66\times10^{-5}$& $1.23$e3& $0.096$& $1.787$&-& $1.403$\\
&    * & $2.66$e-4& $4.15$e3& $6.13\times10^{-3}$& -       & $0.098$& $1.693$& Fe CC-SN& $1.403$\\

\hline\\[-8pt]

$0.158$& $1.9$& $9.36$e-3& $1.11$e5& $9.78\times10^{-6}$& $1.32$e4& $0.251$& $1.11$&-& $1.409$\\
&    * & $1.61$e-4& $2.06$e3& $5.92\times10^{-6}$& $3.98$e3& $0.253$& $1.104$&-& $1.410$\\
&    **& $1.40$e-4& $2.08$e3& $3.06\times10^{-6}$& -       & $0.254$& $1.102$& ONe WD& $1.410$\\

& $2.0$& $7.68$e-3& $9.13$e4& $1.39\times10^{-5}$& $1.20$e4& $0.229$& $1.184$&-& $1.408$\\
&    * & $9.05$e-5& $1.30$e3& $9.86\times10^{-6}$& $4.16$e3& $0.231$& $1.117$&-& $1.408$\\
&    **& $2.20$e-4& $2.73$e3& $1.33\times10^{-5}$& $1.49$e3& $0.236$& $1.159$&-& $1.408$\\
&   ***& $1.38$e-4& $1.80$e3& $8.20\times10^{-6}$&-        & $0.237$& $1.153$&ONe WD& $1.408$\\

& $2.2$& $4.56$e-3& $5.55$e4& $2.19\times10^{-5}$& $8.75$e3& $0.182$& $1.401$&-& $1.405$\\
&    * & $1.73$e-4& $2.26$e3& $3.08\times10^{-5}$& $1.81$e3& $0.188$& $1.359$&-& $1.405$\\
&    **& $1.27$e-4& $1.64$e3& $5.47\times10^{-5}$& $1.98$e2& $0.193$& $1.330$&-& $1.405$\\
&   ***& $8.05$e-5& $9.68$e2& $1.12\times10^{-4}$&-        & $0.200$& $1.288$& ONe WD& $1.405$\\

& $2.4$& $3.77$e-3& $4.58$e4& $3.22\times10^{-5}$& $5.89$e3& $0.159$& $1.537$&-& $1.404$\\
&    * & $2.03$e-4& $3.03$e3& $4.93\times10^{-5}$& $1.06$e3& $0.164$& $1.475$&-& $1.404$\\
&    **& $1.25$e-4& $1.51$e3& $9.27\times10^{-4}$&-        & $0.176$& $1.369$& EC-SN& $1.404$\\

& $2.6$& $3.36$e-3& $4.09$e4& $5.14\times10^{-5}$& $4.43$e3& $0.140$& $1.643$&-& $1.403$\\
&    * & $2.68$e-4& $4.48$e3& $6.44\times10^{-5}$& $8.58$e2& $0.144$& $1.566$&-& $1.404$\\
&    **& $5.63$e-5& $7.40$e2& $3.06\times10^{-3}$&-        & $0.150$& $1.490$& Fe CC-SN& $1.404$\\

& $2.8$& $2.55$e-3& $3.21$e4& $7.08\times10^{-5}$& $3.72$e2& $0.120$& $1.824$&-& $1.403$\\
&     *& $2.78$e-4& $4.40$e3& $3.59\times10^{-3}$&-        & $0.123$& $1.703$&Fe CC-SN& $1.403$\\

\hline\\[-8pt]

$0.200$& $1.9$& $7.59$e-3& $9.12$e4& $1.28\times10^{-5}$& $1.26$e4& $0.312$& $1.122$&-& $1.408$\\
&    * & $1.39$e-4& $1.80$e3& $7.00\times10^{-6}$& $3.96$e3& $0.315$& $1.115$&-& $1.408$\\
&    **& $1.99$e-4& $2.70$e3& $4.92\times10^{-6}$&-        & $0.317$& $1.110$&ONe WD & $1.408$\\

& $2.0$& $6.29$e-3& $7.58$e4& $1.61\times10^{-5}$& $1.12$e4& $0.284$& $1.199$&-& $1.406$\\
&    * & $7.26$e-5& $1.19$e3& $1.16\times10^{-5}$& $3.68$e3& $0.286$& $1.193$&-& $1.406$\\
&    **& $2.26$e-4& $2.78$e3& $1.43\times10^{-5}$& $1.65$e3& $0.293$& $1.172$&-& $1.407$\\
&   ***& $1.76$e-4& $2.19$e3& $1.15\times10^{-5}$&-        & $0.297$& $1.162$& ONe WD& $1.407$\\

& $2.2$& $3.62$e-3& $4.52$e4& $2.71\times10^{-5}$& $8.43$e3& $0.227$& $1.424$&-& $1.404$\\
&    * & $1.78$e-4& $2.39$e3& $3.49\times10^{-5}$& $1.68$e3& $0.235$& $1.376$&-& $1.404$\\
&    **& $2.05$e-4& $2.64$e3& $1.27\times10^{-4}$&-        & $0.251$& $1.296$& ONe WD& $1.404$\\

& $2.4$& $2.99$e-3& $3.74$e4& $3.86\times10^{-5}$& $5.35$e3& $0.198$& $1.563$&-& $1.403$\\
&    * & $2.24$e-4& $3.42$e3& $5.44\times10^{-5}$& $9.67$e2& $0.205$& $1.493$&-& $1.403$\\
&    **& $1.15$e-4& $1.38$e3& $3.72\times10^{-3}$&-        & $0.221$& $1.376$& EC-SN& $1.403$\\

& $2.6$& $2.71$e-3& $3.35$e4& $5.67\times10^{-5}$& $3.67$e3& $0.174$& $1.675$&-& $1.403$\\
&    * & $3.12$e-4& $4.99$e3& $6.86\times10^{-5}$& $7.19$e2& $0.180$& $1.589$&-& $1.403$\\
&    **& $5.28$e-5& $6.96$e2& $3.12\times10^{-2}$&-        & $0.194$& $1.448$& Fe CC-SN& $1.403$\\

\hline\\[-8pt]

$0.251$& $1.9$& $6.23$e-3& $7.53$e4& $1.70\times10^{-5}$& $1.33$e4& $0.389$& $1.132$&-& $1.406$\\
&    * & $1.04$e-4& $1.38$e3& $8.40\times10^{-6}$& $4.03$e3& $0.392$& $1.126$&-& $1.406$\\
&    **& $2.34$e-4& $3.06$e3& $6.28\times10^{-6}$&-        & $0.396$& $1.118$& ONe WD& $1.407$\\

& $2.0$& $5.10$e-3& $6.22$e4& $2.04\times10^{-5}$& $1.05$e4& $0.353$& $1.214$&-& $1.405$\\
&    * & $7.68$e-5& $1.19$e3& $1.22\times10^{-5}$& $3.11$e3& $0.355$& $1.208$&-& $1.405$\\
&    **& $2.34$e-4& $2.89$e3& $1.52\times10^{-5}$& $1.79$e3& $0.365$& $1.183$&-& $1.405$\\
&   ***& $1.92$e-4& $2.36$e3& $1.41\times10^{-5}$&-        & $0.372$& $1.168$&-& $1.406$\\

& $2.2$& $2.90$e-3& $3.67$e4& $3.51\times10^{-5}$& $8.20$e3& $0.283$& $1.442$&-& $1.403$\\
&    * & $1.84$e-4& $2.49$e3& $3.89\times10^{-5}$& $1.60$e3& $0.293$& $1.387$&-& $1.403$\\
&    **& $2.07$e-4& $2.55$e3& $1.51\times10^{-4}$&-        & $0.315$& $1.301$& ONe WD& $1.403$\\


& $2.3$& $3.49$e-3& $4.34$e4& $3.67\times10^{-5}$& $1.02$e4& $0.279$& $1.422$&-& $1.403$\\
&    * & $1.79$e-4& $2.37$e3& $3.57\times10^{-5}$& $2.09$e3& $0.289$& $1.372$&-& $1.404$\\
&    **& $1.34$e-4& $1.74$e3& $5.82\times10^{-5}$& $2.61$e2& $0.297$& $1.339$&-& $1.404$\\
&   ***& $7.48$e-5& $8.97$e2& $1.35\times10^{-4}$&-        & $0.310$& $1.291$& ONe WD& $1.404$\\

\hline\\[-8pt]

$0.316$& $1.9$& $5.22$e-3& $6.30$e4& $2.16\times10^{-5}$& $1.28$e4& $0.484$& $1.141$&-& $1.405$\\
&    * & $1.05$e-4& $1.42$e3& $9.10\times10^{-6}$& $4.03$e3& $0.489$& $1.133$&-& $1.405$\\
&    **& $2.67$e-4& $3.41$e3& $7.49\times10^{-6}$&-     & $0.496$& $1.122$& ONe WD& $1.406$\\

& $2.0$& $4.18$e-3& $5.12$e4& $2.60\times10^{-5}$& $1.00$e4& $0.439$& $1.226$&-& $1.404$\\
&    * & $9.27$e-5& $1.41$e3& $1.39\times10^{-5}$& $2.82$e3& $0.442$& $1.218$&-& $1.404$\\
&    **& $2.30$e-4& $2.85$e3& $1.63\times10^{-5}$& $1.76$e3& $0.456$& $1.191$&-& $1.404$\\
&   ***& $1.95$e-4& $2.37$e3& $1.62\times10^{-5}$&-        & $0.465$& $1.174$&-& $1.405$\\

& $2.1$& $3.02$e-3& $3.77$e4& $3.51\times10^{-5}$& $1.05$e4& $0.388$& $1.348$&-& $1.403$\\
&    * & $1.74$e-4& $2.30$e3& $2.51\times10^{-5}$& $2.53$e3& $0.399$& $1.313$&-& $1.403$\\
&    **& $1.54$e-4& $1.98$e3& $4.04\times10^{-5}$& $6.42$e2& $0.411$& $1.282$&-& $1.403$\\
&   ***& $1.45$e-4& $1.72$e3& $5.15\times10^{-5}$&-        & $0.426$& $1.245$&ONe WD& $1.403$\\

\hline\\[-8pt]

$0.398$& $1.8$& $5.57$e-3& $6.65$e4& $2.06\times10^{-5}$& $1.45$e4& $0.652$& $1.085$&-& $1.406$\\
&    * & $1.48$e-4& $2.45$e3& $1.66\times10^{-6}$&-& $0.653$& $1.083$& ONe WD& $1.406$\\

& $2.0$& $3.50$e-3& $4.31$e4& $3.30\times10^{-5}$& $9.75$e3& $0.547$& $1.235$&-& $1.403$\\
&    * & $1.05$e-4& $1.53$e3& $1.15\times10^{-5}$& $2.73$e3& $0.553$& $1.226$&-& $1.404$\\
&    **& $2.25$e-4& $2.78$e3& $1.83\times10^{-5}$& $1.75$e3& $0.571$& $1.197$&-& $1.404$\\
&   ***& $1.92$e-4& $2.29$e3& $1.79\times10^{-5}$&-        & $0.583$& $1.177$&ONe WD& $1.404$\\

& $2.1$& $2.51$e-3& $3.16$e4& $4.20\times10^{-5}$& $1.04$e4& $0.484$& $1.359$&-& $1.403$\\
&    * & $1.75$e-4& $2.31$e3& $2.72\times10^{-5}$& $2.49$e3& $0.500$& $1.320$&-& $1.403$\\
&    **& $1.52$e-4& $1.96$e3& $4.45\times10^{-5}$& $5.56$e2& $0.515$& $1.288$&-& $1.403$\\
&   ***& $1.40$e-4& $1.65$e3& $5.67\times10^{-5}$&-        & $0.535$& $1.248$& ONe WD& $1.403$\\

\hline\\[-8pt]

$0.501$& $1.8$& $4.89$e-3& $5.90$e4& $2.45\times10^{-5}$& $1.50$e4& $0.816$& $1.089$&-& $1.405$\\
&    * & $1.91$e-4& $2.87$e3& $2.29\times10^{-6}$&-        & $0.819$& $1.086$&ONe WD& $1.405$\\

& $2.0$& $2.96$e-3& $3.70$e4& $3.92\times10^{-5}$& $9.52$e3& $0.684$& $1.243$&-& $1.403$\\
&    * & $8.84$e-5& $1.37$e3& $1.32\times10^{-5}$& $2.58$e3& $0.690$& $1.234$&-& $1.403$\\
&    **& $2.33$e-4& $2.85$e3& $1.92\times10^{-5}$& $1.73$e3& $0.715$& $1.201$&-& $1.403$\\
&   ***& $1.84$e-4& $2.17$e3& $1.90\times10^{-5}$& $8.29$e2& $0.732$& $1.180$&-& $1.403$\\
&  ****& $1.50$e-5& $2.70$e2& $1.10\times10^{-6}$&-        & $0.733$& $1.180$&ONe WD& $1.403$\\

\hline\\[-8pt]

$0.631$& $1.8$& $4.33$e-3& $5.20$e4& $2.86\times10^{-5}$& $1.44$e4& $1.020$& $1.093$&-& $1.404$\\
&    * & $2.42$e-4& $3.36$e3& $3.25\times10^{-6}$&-        & $1.030$& $1.089$& ONe WD& $1.405$\\

& $2.0$& $2.53$e-3& $3.21$e4& $4.76\times10^{-5}$& $9.33$e3& $0.855$& $1.250$&-& $1.403$\\
&    * & $9.53$e-5& $1.44$e3& $1.45\times10^{-5}$& $2.47$e3& $0.863$& $1.240$&-& $1.403$\\
&    **& $2.38$e-4& $2.94$e3& $2.02\times10^{-5}$& $1.61$e3& $0.896$& $1.205$&-& $1.403$\\
&   ***& $1.75$e-4& $2.06$e3& $2.00\times10^{-5}$& $6.99$e2& $0.919$& $1.183$&-& $1.403$\\
&  ****& $2.94$e-5& $4.15$e2& $3.03\times10^{-6}$&-        & $0.920$& $1.182$& ONe WD& $1.403$\\

\hline\\[-8pt]

$0.794$& $1.8$& $3.88$e-3& $4.69$e4& $3.33\times10^{-5}$& $1.37$e4& $1.280$& $1.096$&-& $1.404$\\
&    * & $2.61$e4& $3.59$e3& $3.76\times10^{-6}$& $1.290$& $1.090$& ONe WD& $1.404$\\

& $1.9$& $3.01$e-3& $3.72$e4& $4.22\times10^{-5}$& $1.19$e4& $1.190$& $1.161$&-& $1.403$\\
&    * & $1.11$e-4& $1.49$e3& $1.15\times10^{-5}$& $3.66$e3& $1.210$& $1.150$& ONe WD& $1.403$\\
&    **& $3.00$e-4& $3.69$e3& $1.03\times10^{-5}$&-        & $1.230$& $1.132$& ONe WD& $1.403$\\

\hline\\[-8pt]

  $1.0$& $1.8$& $3.49$e-3& $4.25$e4& $3.79\times10^{-5}$& $1.35$e4& $1.610$& $1.098$&-& $1.403$\\
  &     *& $2.77$e-4& $3.69$e3& $4.24\times10^{-6}$&-& $1.620$& $1.092$&ONe WD& $1.404$\\
  
  & $1.9$& $2.69$e-3& $3.30$e4& $4.90\times10^{-5}$& $1.18$e4& $1.490$& $1.164$&-& $1.403$\\
  &    * & $1.15$e-4& $1.50$e3& $1.21\times10^{-5}$& $3.63$e3& $1.510$& $1.152$&-& $1.403$\\
  &    **& $3.06$e-4& $3.72$e3& $1.08\times10^{-5}$&-        & $1.550$& $1.133$& ONe WD& $1.403$\\
  
\hline\\[-8pt]
  
$1.259$& $1.8$& $3.17$e-3& $3.88$e4& $4.30\times10^{-5}$& $1.31$e4& $2.020$& $1.101$&-& $1.403$\\
       &     *& $2.89$e-4& $3.74$e3& $4.63\times10^{-6}$&-        & $2.040$& $1.093$& ONe WD& $1.403$\\
       

\hline\\[-8pt]

$1.585$& $1.8$& $2.89$e-3& $3.49$e4& $4.90\times10^{-5}$& $1.29$e4& $2.530$& $1.103$&-& $1.403$\\
       &    * & $2.92$e-4& $3.48$e3& $5.22\times10^{-6}$&-        & $2.560$& $1.094$& ONe WD& $1.403$\\
       
\hline\\[-8pt]

$1.995$& $1.8$& $2.65$e-3& $3.17$e4& $5.52\times10^{-5}$& $1.27$e4& $3.170$& $1.106$&-& $1.403$\\
       &    * & $3.05$e-4& $3.63$e3& $5.79\times10^{-6}$&-        & $3.220$& $1.095$& ONe WD& $1.403$\\
\end{longtable}




\label{lastpage}
\end{document}